# MAGNETO INERTIAL FUSION BASED ON A CUSP FIELD CONFIGURATION

S.V. Ryzhkov[1], I.Yu. Kostyukov[2]

[1]*Bauman Moscow State Technical University, Moscow, Russia*
[2]*Institute of Applied Physics RAS, Nizhny Novgorod, Russia*

The magneto inertial fusion (MIF) is the fastest developing area of science, which combines the advantages of traditional approaches to the fusion: magnetic confinement fusion and inertial confinement fusion [1]. The first MIF experiments based on laser-driven magnetic-flux compression have been carried out recently [2]. Magnetized cylindrical targets were imploded on OMEGA laser system to compress a pre-seeded magnetic flux to multi-megagauss values. The experiments demonstrated the bright prospects of laser-driven inertial confinement fusion with magnetized targets.

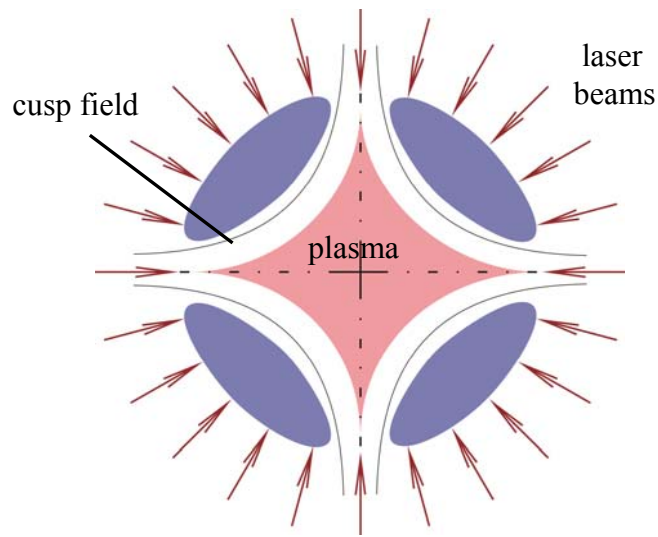

Fig. 1. The implosion scheme – the spherical target (plasma of cusp magnetic field configuration) compression under laser beams.

The configuration of pre-seeded magnetic field used in the experiments [2] is close to solenoid-like, which is far from optimal because of large particle losses from the solenoid ends. We propose to use the cusp configuration of the pre-seeded magnetic field (see Fig. 1). The cusp configuration provides better confinement of the fast charged particles than the configuration used in [2]. It also provides implosion that close to spherical. Like in the experiments [2] the pre-seeded magnetic can be easily generated by two oppositely directed high-current loops. The proposed magnetic field distribution and the particle trajectories in the filed are shown in Fig. 2.

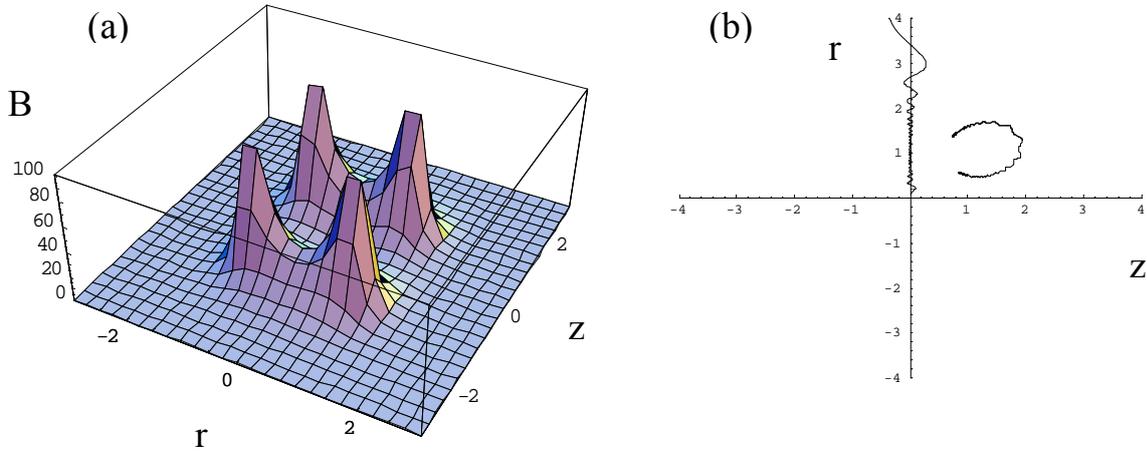

Fig. 2. The distribution of the magnetic field intensity (a); two examples of the particle trajectory (trapped and untrapped particles) in the magnetic field of two oppositely directed current loops calculated by numerical integration of equation of motion (b).

Following Ref. [3] we develop the model describing the plasma implosion and particle losses from magnetic trap. We assume that the plasma inside can be represented by the equivalent sphere of radius $r$ and is confined by the spherical shell of mass $m$ and radius $r$. The shell implosion can be described by the equation [4]

$$\frac{d^2 r}{dt^2} = \frac{4\pi r^2 p}{m}, \qquad (1)$$

where $p$ is the shell pressure, $m$ is the total fuel mass $m = 4\pi \rho_{in} r_{in}^2 \Delta r_{in}$, where $r_{in}$ is the initial radius (the capsule radius) and $\Delta r_{in}$ is the thickness of the shell. Making of the use

$$NkT = pV, \qquad (2)$$

Eq. (1) may be written

$$\frac{d^2 r}{dt^2} = \frac{3NT}{rm}. \qquad (3)$$

The particle loss from cusp is given by equation [5]

$$\frac{dN}{dt} = -\frac{16nTrc}{eB}, \qquad (4)$$

where $n$ is the plasma density, $T$ is the plasma temperature, $B$ is magnetic field value. The magnetic filed compression can be described by relation [2]

$$B \approx B_0 \left(\frac{r_0}{r}\right)^{2(1-1/\mu)}, \qquad (5)$$

where μ is the magnetic Reynolds number. Below we assume that $\mu \gg 1$. Combining Eqs. (4), (5) we derive equations

$$\frac{dN}{dt} = -\frac{N}{\tau_{loss}}, \tag{6}$$

$$\tau_{loss} = \frac{\pi}{12} \omega_{ci} \frac{r_0^2}{v_i^2}, \tag{7}$$

where $r_{in} = r(t = t_0)$ is the initial radius of sphere, $v_i = (T_i / m_i)^{1/2}$ is the ion thermal velocity, $T_i$ is the plasma temperature. Solution of the Eq. (6) is:

$$\tilde{N} = \frac{N}{N_0} = \exp(-\frac{t}{\tau_{loss}}), \tag{8}$$

where $N_0 = N(t = t_0)$ is the initial number of plasma particles.

Substituting solution (8) into Eq. (4), we get the equation describing the compression of the shell

$$\frac{d^2 a}{d\tau^2} = \frac{1}{a} \exp[-\alpha \tau], \tag{9}$$

where $a = r/r_0$, $\tau = t/\tau_{comp}$ are dimensionless parameters, $\tau_{comp} = (4\pi p_{in} / m)^{-1/2}$ is the compression time, $\alpha = \tau_{comp} / \tau_{loss}$, $r_0$ is the minimal radius of the plasma, $p_{in}$ is the initial pressure on the shell.

We numerically solve system of equations (8) and (9) for initial solution $a(t=0) = 1$ and $\dot{a}(t=0) = 0$ for different values $\alpha$. The solutions of Eq. (9) and magnetic field normalized on maximum field $\tilde{B} = B/B_0$ are shown in Fig. 3.

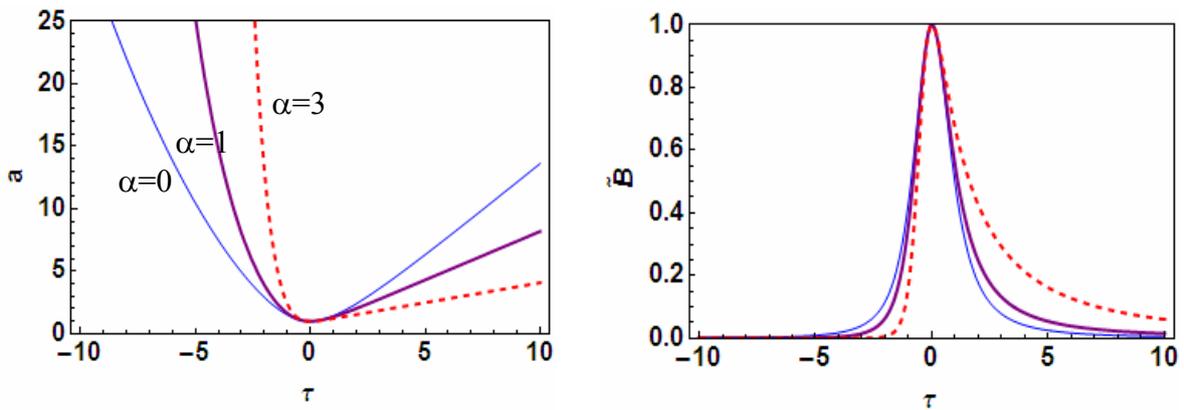

Fig. 3. Dimensionless radius ($a$) and magnetic field ($\tilde{B}$) distributions versus dimensionless time ($\tau$) for spherical cusp configuration for different values.

In Conclusion we study the cusp configuration of pre-seeded magnetic field for laser-driven flux-compression. The proposed configuration provides better particle confinement than that in Ref. [2] thereby leading to better magnetic insulation of hot fusion plasma. We develop the model that takes into account (i) magnetized target implosion, (2) magnetic flux compression; (iii) particle losses from magnetic trap. The key parameter governing the process is $\alpha$ which is the ratio of the implosion time to the particle loss time. If $\alpha \gg 1$ than the particle losses are significant. In the opposite limit the losses are not large, which is more favorable for fusion. The resulting magnetic field should be high enough for regime with $\alpha \ll 1$. It should be noted that additional electrodes with high ambipolar potential located near cusp can significantly suppresses particle losses from magnetic tarp.

The work is supported by the Russian Foundation for Basic Research (RFBR Grants № 09-02-90702-mob_st, № 09-08-00137-a, № 07-02-01239-a).

References
   [1] I.R. Lindemuth, R.C. Kirkpatrick, Nuclear Fusion 23, 263-284 (1983)
   [2] O.V. Gotchev, N.W. Jang, J.P. Knauer, M.D. Barbero, R. Betti, C.K. Li, R.D. Petrasso, Journal of Fusion Energy 27, 25-31 (2008)
   [3] A.E. Robson, Naval Research Laboratory Report MR-2692 (1973)
   [4] S. Pfalzner, An Introduction to Inertial Confinement Fusion. Series in Plasma Physics. Taylor & Francis: New York, London (2006)
   [5] I.J. Spalding, Nuclear Fusion 8, 161 (1968)